\documentclass[11pt]{article}
\usepackage{graphicx}
\usepackage{amsmath}
\usepackage[a4paper, total={6in, 8in}]{geometry}

\title{\bf Effect of Oxygen Vacancy Defects on Electronic and Optical Properties of MgO Monolayers: First Principles Study}
\author{ Rituparna Hazarika, Bulumoni Kalita\footnote{Corresponding Author, bulumonikalita@dibru.ac.in}\\
Dept. of Physics, Dibrugarh University,Assam, India, 786001}

\begin{document}
\maketitle

\begin{abstract}
The optoelectronic properties induced by oxygen vacancy defects in MgO(111) monolayers have been studied using hybrid level of DFT method.  HSE calculations shows significant reduction in electronic band gap of MgO monolayer as a result of introduction of oxygen vacancies.  The pristine monolayer has a wide band gap (4.84 eV, indirect) semiconducting behaviour, which changes gap to 2.97 eV (indirect) and 2.28 eV (direct) with increment in oxygen vacancy defect concentration of 6.25\% and 12.5\%, respectively.  Consequently, presence of oxygen vacancies leads to energy red shift of the observed optical phenomena with reference to the pristine monolayer.  Most importantly, the divacancy system with two consecutive vacancy sites displays the strongest optical absorption and also becomes optically responsive over the spectral range from visible to ultraviolet region in the electromagnetic spectrum.  Thus creation of oxygen vancancies in MgO monolayers may be a fruitful technique for achieving a suitable solar energy material.
\end{abstract}

\section{Introduction}\label{sec1}
Graphene-like two-dimensional (2D) materials have gained tremendous research interest over the last two decades due to their fascinating electronic, optical, mechanical properties \cite{bib1}.  Motivated by these facts, a huge group of researchers have performed both theoretical and experimental studies on a numerous single layered structures such as transition metal dichalcogenides (TMDCs) \cite{bib2} and trichalcogenides (TMTs) \cite{bib3}; elemental and binary monolayers (MLs) based on group IV and V \cite{bib4}; binary MLs of group II-VI \cite{bib5}; group III-V \cite{bib6}; group III-VI \cite{bib7}; group IV-VI \cite{bib8} etc.  Further, electronic structure modification in 2D materials through various techniques leads to enhancement in their properties \cite{bib1, bib9}.  As a result, the large variety of 2D materials finds potential applications in numerous fields, viz., catalysis, optoelectronics and sensing devices, energy storage and conversion etc. In a latest theoretical study, honeycomb-structured MLs of a group of monoxides and monochlorides have been found to possess high dynamic, kinetic, mechanical and thermal stabilities \cite{bib10}.  Another study based on 32 II-VI hexagonal single layer structures has revealed that BeO, MgO, CaO, ZnO, CdO MLs depict good dynamical stability \cite{bib11}.
\paragraph {}
Bulk magnesium oxide (MgO) is a highly insulating material with a wide band gap of 7.8 eV \cite{bib12}.  This non-magnetic oxide is technologically important because of its simple rock salt structure and non-toxicity.  There is significant reduction in the band gaps of MgO in their nansotructure forms \cite{bib13, bib14}.  Moreover, applicability of these materials in diverse areas such as ultra-violet (UV) photodetector, spintronics, catalysis, waste water treatment, antibacterial activities, biomedical applications, clean energy production etc. make them more interesting \cite{bib15}.  Among the different types of MgO nanostructures, immense efforts have been rendered to synthesize and study two-dimensional MgO nanosheets.  Both polar (111) and non-polar (100) facets of 2D MgO thin films have been successfully synthesized on various substrates \cite{bib16, bib17}.  K. Matsuzaki \emph{et al.} had grown graphene-like MgO layers epitaxially on yttrium-zirconia (111) substrate and revealed the height of MgO single layer to be 2.43\AA \cite{bib16}.  Experimental study performed by S. Benedetti \emph{et al.} demonstrated that distinguished experimental conditions such as deposition temperature, oxygen partial pressure etc. are necessary for the growth of MgO(111) and MgO(100) thin films \cite{bib17}.  The (111) MgO ML exhibits pure graphene like planar hexagonal structure as observed from experimental and theoretical studies \cite{bib16, bib18} and also it is found to be mechanically stable in contrast with the (100) ML \cite{bib18}.  It has also been observed that MgO monolayers show semiconducting nature and the band gaps tuning of MgO nanobelts is also possible by changing their widths \cite{bib18}.  Improvements in various properties of MgO nanomaterials are needed for device applications and this can be achieved through different engineering techniques.  They are like structural modification by adding dopants, functionalization, creating various defects, applying stress/strain etc.
\paragraph {}
It has been observed theoretically that doping with light 2p elements and transition metals introduces magnetic behaviour in MgO (111) MLs \cite{bib19, bib20}.  Previous density functional theory (DFT) study showed that MgO (111) MLs have high transmittance and low reflectivity for a wide energy range \cite{bib21}.  Later, another DFT work has portrayed the honeycomb-like MgO ML as a potential candidate for optoelectronic applications \cite{bib22}.  M. Yeganeh and F. Kafi investigated the variation in optical properties of graphene like MgO (111) MLs under the application of strain \cite{bib23}.  N and F functionalizations are found to enhance the optical properties of MgO single layer over the pristine \cite{bib24}.  Numbers of experimental and theoretical investigations have addressed the effect of vacancies in the optical properties of MgO bulk, surface and other nanoforms \cite{bib25, bib26, bib27, bib28}.  Presence of F-centres in bulk MgO is found to increase the number of optical absorption lines \cite{bib29} and also it affects both the absorption and emission spectra \cite{bib30}.  MgO surfaces with periodic F defects have been reported to display non-linear optical properties \cite{bib31}.  Enhancement in transmittance (80\%) in IR part of electromagnetic spectrum is observed for anion deficient nanocrystalline MgO powder \cite{bib25}.  Similarly, vacancy defects affecting the optical properties in different other monoxide monolayers have also been studied very recently.  It has been found that ZnO ML with oxygen vacancy in presence of strain \cite{bib32} and Yttrium dopant \cite{bib33} leads to increment in solar light region absorption coefficient compared to that of its pristine monolayer.  BeO ML induced by vacancy and anti-site defects exhibit better optoelectronic performance \cite{bib34}.  D. M. Hoat \emph{et al.} has concluded that depletion in energy gap of BeO ML with oxygen vacancies may improve its optical properties \cite{bib35}.
\paragraph {}
From the numerous available literatures, it is clear that presence of oxygen vacancy can greatly improve the optical properties of 2D graphene like oxide monolayers.  In spite of this, studies on MgO(111) MLs with vacancy defects are still missing to the best of our knowledge.  Motivated by these facts, in the present work we will theoretically investigate the optical properties of MgO(111) MLs containing oxygen vacancy defects and compare them with those of the pristine monolayer.
\section {Computational Details}
The calculations are performed using density functional theory (DFT) methods as implemented in the freely available Quantum ESPRESSO (QE) software \cite{bib36}.  QE code uses a plane-wave basis set to solve Kohn-Sham equations \cite{bib37}.  The interactions of valence electrons and ions are sketched with projector augmented wave (PAW) pseudopotential \cite{bib38} and Perdew-Burke-Ernzerhof (PBE) form of generalized gradient approximation (GGA) \cite{bib39} depicts the exchange-correlation part.  All the calculations based on the PAW approach used energy cut off value of 30 Ry and 240 Ry for wave functions and charge densities, respectively, which are obtained by convergence tests.  Further, it is well known that the GGA-PBE functional underestimates the energy gap of a system \cite{bib40}.  To overcome this limitation, hybrid functional of Heyd, Scuseria and Ernzerhof (HSE) \cite{bib41} functional has already been effectively applied earlier \cite{bib10, bib11, bib24, bib42}.  We have therefore used HSE functional with optimized norm-conserving pseudopotentials (ONCV) \cite{bib43} to achieve accurate electronic band structure of our considered systems.  For hybrid calculations, the wave functions and charge density cut off values are chosen to be 70 Ry and 280 Ry respectively.  We have constructed a $4\times4\times1$ supercell for MgO (111) monolayer containing 32 atoms, where 16 Mg atoms and 16 O atoms are present.  The modelled MgO supercell has regular pattern along x, y directions and 15 Å vacuum is introduced to disturb the periodicity in z-direction.  To study the impact of oxygen vacancy defects on MgO monolayer, we have built two supercells.  In the first, we have removed one O atom that corresponds to defect concentration of 6.25\% and in the second, we have removed of two O atoms simultaneously resulting in defect concentration of 12.5\%.  A  $2\times2\times1$ Monkhorst-Pack k-point grid \cite{bib44} is adapted for sampling the Brillouin zone of the MgO monolayers.  We have carried out spin- polarized calculations and van der Waals (vdW) interactions are treated by Grimme’s D2 method \cite{bib45} for all the MgO monolayers.  The geometries of the considered systems are relaxed until the total energy and force convergence threshold of 10-5 Ry and 10-3 Ry/Bohr, respectively, are met.  Marzari–Vanderbilt smearing \cite{bib46} of width 0.002 Ry is utilized for smooth conduction of convergence. 
\paragraph {}
In order to calculate the stability of the pristine and defective MgO monolayers,we have calculated the binding energy per atom given by the formula,
\begin{equation}
BE/atom = \frac{E_{tot}-(N_{Mg}E_{Mg}+N_{O}E_{O})}{(N_{Mg}+N_{O})}
\end{equation}
Where,$ E_{tot}$ is the total energy of the MgO monolayer system.  $N_{Mg}$, $N_{O}$ are the number of Mg and O atoms in the supercell considered and $E_{M}g$, $E_{O}$ are the energies of individual Mg and O atoms, respectively.
Next, we have calculated the vacancy formation energy given by the formula, 
\begin{equation}
E_{Vf} = E_{V}+E_{O}-E_{P}
\end{equation}
Where $E_{V}$ is the total energy of the MgO monolayer with single and double O vacancies, $E_{O}$ is the energy of the individual O atom, $E_{P}$ is the total energy of the pristine MgO monolayer.
HSE functional has also been employed to compute the optical properties using Epsilon.x code, which is incorporated with random phase approximation (RPA) \cite{bib47}.  In this approximation, the imaginary part $(\epsilon_{i}(\omega))$ of the frequency dependent complex dielectric function $(\epsilon(\omega)=\epsilon_{r}(\omega)+i\epsilon_{i}(\omega))$ is described by perturbation theory in adiabatic conditions and is represented by the expression
\begin{equation}
\epsilon_{i,\alpha\beta}(\omega) = \frac{4{\pi^2}{e^2}}{\Omega}\lim\limits_{q \to0}\frac{1}{q^2}\sum_{c,v,k}2w_{k}\delta(\epsilon_{ck}-\epsilon_{vk}-\omega)<U_{ck+e_{\alpha}q}|{U_{vk}}><U_{ck+e_{\beta}q}|U_{vk}>
\end{equation}
Where $\omega$ is the frequency of electromagnetic radiation in energy units,  is the volume of the primitive cell, indices c, v and k represents conduction band, valence band and reciprocal space point respectively, $w_{k}$ represents the k-point weights, $U_{ck}$ is the cell periodic part of the orbitals at the k point, $e_{\alpha}$ is the unit vector for the three cartesian directions.  The real part  $(\epsilon_{r}(\omega))$ can be calculated using Kramers-Kronig relation \cite{bib48, bib49, bib50} from the imaginary part  $(\epsilon_{i}(\omega))$ and is expressed as
\begin{equation}
\epsilon_{r,\alpha\beta}(\omega) = 1+\frac{2}{\pi}P\int_{0}^{\infty}\frac{\epsilon_{i,\alpha\beta}(\omega^\prime)(\omega^\prime)}{\omega^{\prime2}-\omega^2}d\omega^\prime
\end{equation}
Where P is the principle value of integral.
From the computed values of real and imaginary parts of dielectric function, the other frequency dependent optical properties such as refractive index$(\epsilon_{i}(\omega))$, extinction coefficient $(k(\omega))$, reflectivity $(R(\omega))$, absorption coefficient $(\alpha(\omega))$ and electron energy loss function $(L(\omega))$ are determined by applying the following formulae –
\begin{equation}
n(\omega) = \frac{1}{\sqrt{2}}[(\epsilon_{r}(\omega)^{2}+\epsilon_{i}(\omega)^2)^{\frac{1}{2}}+\epsilon_{r}(\omega)]^{\frac{1}{2}}
\end{equation}
\begin{equation}
k(\omega) = \frac{1}{\sqrt{2}}[(\epsilon_{r}(\omega)^{2}+\epsilon_{i}(\omega)^2)^{\frac{1}{2}}-\epsilon_{r}(\omega)]^{\frac{1}{2}}
\end{equation}
\begin{equation}
R(\omega) = \frac{(n(\omega)-1)^{2}+k(\omega)^{2}}{(n(\omega)+1)^{2}+k(\omega)^{2}}
\end{equation}
\begin{equation}
\alpha(\omega) = \sqrt{2}\omega[(\epsilon_{r}(\omega)^{2}+\epsilon_{i}(\omega)^2)^{\frac{1}{2}}-\epsilon_{r}(\omega)]^{\frac{1}{2}}
\end{equation}
\begin{equation}
\label{simple equation}
L(\omega) = \frac{\epsilon_{i}(\omega)}{\epsilon_{i}(\omega)^{2}+\epsilon_{r}(\omega)^{2}}
\end{equation}
\section {Results and Discussions}
\subsection {Structure and stability}
The optimized $4\times4\times1$ supercells of pristine (P) and defective MgO (111) monolayers are shown in Figure 1.  The formation energy along with structural and stability parameters of the relaxed structures is tabulated in Table 1.  Defective MgO ML with single O vacancy defect is represented as $V_{O}$, whereas there are five different types are available for the double O vacancy layer.   They are constructed by choosing the vacancy sites as i) A-B ($V_{2O}$-I) ii) A-C ($V_{2O}$-II) iii) A-D ($V_{2O}$-III), iv) A-E ($V_{2O}$-IV), v) A-F ($V_{2O}$-V) (Figures 1 and S1 in the Supplementary Information).  Positive values of formation energy represent endothermic formation of the oxygen vacancies.  The $E_{Vf}$ values of the double vacancy MgO MLs are almost similar and they are almost double that of the monovacancy MgO ML.  The negative BE/atom values indicate that all the MgO monolayers are stable with the pristine being the most stable.  The stability of P has already been confirmed with phonon dispersion analysis in a previous study \cite{bib18, bib24, bib51, bib52}.  The binding energy of this monolayer is less than that of MgO bulk experimental value (10.26 eV) \cite{bib53}.  The stability slowly decreases with the number of defects in the pristine MgO ML.  It is observed that for the divacancy MgO MLs, the formation energy as well as the binding energy does not vary for the structures with the two O vacancies sitting distant apart.  On the other hand, the $V_{2O}$-I MgO monolayer having the O vacancy defects at two consecutive positions shows slightly different $E_{vf}$ than the others.  To mark this distinction, we have chosen only $V_{2O}$-I and $V_{2O}$-II for further studies.  The remaining structures are shown in Figure S1.
\begin{table}
\begin{center}
\caption {Structural and stability parameters of MgO (111) monolayers.}\label {<table-label>}
\vspace{.51cm}
\begin{tabular} {| c|c|c|c| }
\hline
\bf Systems & \bf Position (F centre) & $ \bf E_{vf}$ \bf (eV) & \bf BE/atom (eV)\\
\hline
\bf P & - & - & -4.63\\
\hline
$ \bf V_{O}$ & A & 8.54 & -4.51 \\
\hline
$ \bf V_{2O}$ \bf -I & A-B & 17.14 & -4.37 \\
\hline
$ \bf V_{2O}$ \bf -II & A-C & 17.10 & -4.37 \\
\hline
$ \bf V_{2O}$ \bf -III & A-D & 17.10 & -4.37 \\
\hline
$ \bf V_{2O}$ \bf -IV & A-E & 17.10 & -4.37 \\
\hline
$ \bf V_{2O}$ \bf -V & A-E & 17.10 & -4.37 \\
\hline
\end{tabular}
\end{center}
\end{table}
\paragraph {}
It is found that the relaxed structure of MgO monolayer is planar with lattice constant 3.28\AA and Mg-O bond length 1.90\AA, which matches with previous studies confirming decrement of these values from those of MgO bulk structure .  All the defective monolayers also remain planar with Mg-Mg average distance near the defect site is 3.18\AA, which is comparatively less than the lattice constant of pristine (Figure 1).
\clearpage
\begin{figure}
\centering
\includegraphics[width=0.79\textwidth]{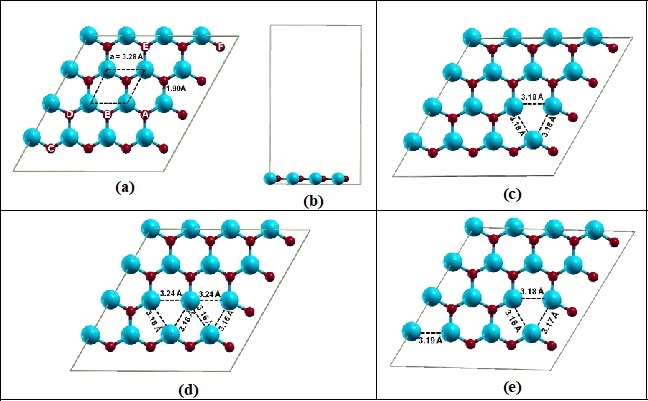}
\caption{Optimized structure of pristine (P) and defective MgO (111) monolayers (a) P - top view (b) P - side view, (c) $V_{O}$ (d) $V_{2O}$-I and (e) $V_{2O}$-II where blue and and red balls represent Mg and O atoms, respectively.}\label{Figure_1}
\end{figure}
\subsection {Electronic properties}
We have computed the band structure, density of states (DOS) of the considered monolayers using both PAW and HSE functionals.  Since HSE functional is capable of describing a more reliable band gap, the band structure and DOS plots using HSE functional are only shown here (Figure 2).  However, the corresponding PAW level computed band structures are shown in Figure S2 for reference.  The Fermi energy levels in the figures, represented by dotted lines are set at zero energy.  For pristine MgO monolayer, the band gap values obtained with PAW and HSE functional are 3.39 eV and 4.84 eV, respectively, indicating it as a wide band-gap semiconductor as reported earlier \cite{bib10,bib11,bib24,bib42}. 
\begin{figure}
\centering
\includegraphics[width=0.9\textwidth]{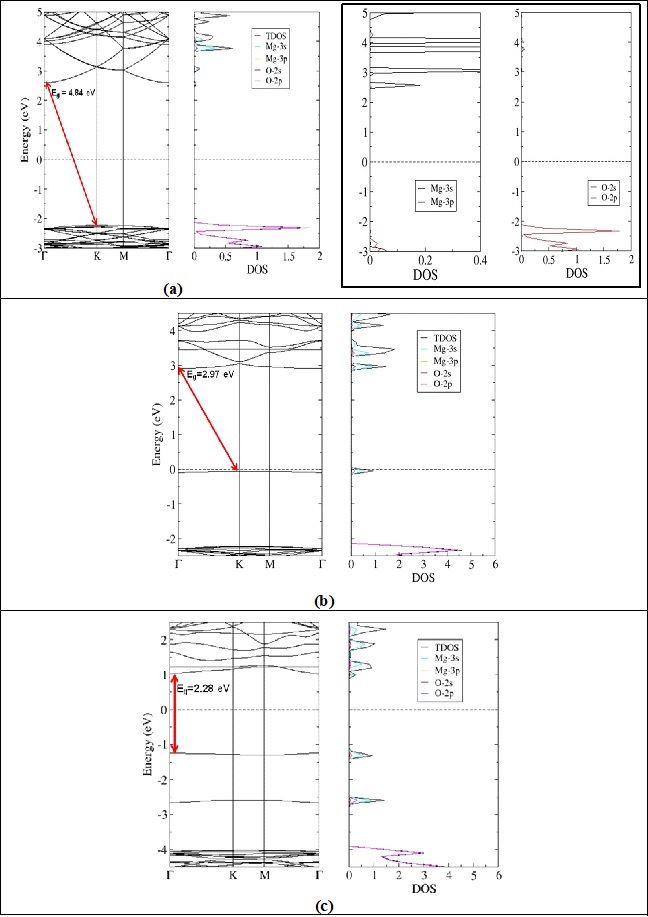}
\caption{Band structures and partial density of states (PDOS) plots of MgO (111) monolayers (a) P (b) $V_{O}$ (c) $V_{2O}$-I (d) $V_{2O}$–II.  The black dotted line drawn perpendicular to the energy axis is Fermi level chosen at 0 eV.}\label{Figure_2}
\end{figure}
\newpage
It is to be noted that P is non-magnetic and its band gap is of indirect nature as shown by both the functionals.  Figure 2(a) shows that the valence band maximum (VBM) appears at K symmetric point along with the conduction band minimum (CBM) at $\gamma$ symmetric point for the pristine MgO monolayer.  Our observations agree well with the previous theoretical results \cite{bib19,bib24}.  All the considered defective monolayers bear non-magnetic characters as previously seen for other oxide monolayers \cite{bib34,bib54}.  The O vacancies introduce defect levels into the band gap of defective MgO monolayers, which result in the decrement of their electronic band gaps.  The indirect band gap nature of P is retained in $V_{O}$ with a decrement in band gap to 2.97eV (Figure 2(b)).  In $V_{O}$, an occupied defect level occurs at 0.059 eV just below the Fermi level.  A similar effect was also oberserved for other nanosheets such as BeO and ZnO \cite{bib32,bib33,bib34,bib35}.  From Figure 2(c), it is seen that $V_{2O}$-I has a direct band gap (at high symmetry $\gamma$ point) with band gap value of 2.28 eV, while $V_{2O}$-II exhibits an indirect band gap value of 2.87 eV with its VBM at $\gamma$ point and CBM at K point (Figure 2(d)).  In case of these divacancy monolayers, two defect levels appear at energies -1.25 eV, -2.59 eV for $V_{2O}$-I and at energies -0.46 eV, -0.73 eV for $V_{2O}$–II.  The relative positions of the Fermi levels and the defect states (Figure 2) signify the defective MgO monolayers to be of n-type semiconducting nature.  Previous studies have shown similar behaviour for other oxide monolayers \cite{bib32,bib33,bib34,bib35}.
\paragraph{}
The DOS analyses are performed to understand the orbital contribution to the valence and the defect states introduced by the vacancies in MgO monolayers.  The total DOS (TDOS) and partial DOS (PDOS) plots are also presented in Figure 2.  From Figure 2(a), it is observed that the VBM and the CBM for pristine MgO monolayer arise from the O (2p) and Mg (3s) states, respectively, which is in line with previous theoretical observations \cite{bib19,bib24}.  Moreover, mixing of s and p orbitals of Mg and O atoms are also visible in the inset in Figure 2(a), which suggests hybridization of such orbitals forming sp2 hybridized states.  The PDOS plots of MgO monolayer with monovacancy (Figure 2(b)) and divacancies (Figures 2(c) and 2 (d)) show that the impurity levels below the Fermi level are mainly contributed by almost all the valence orbitals of Mg and O atoms except O (2s) states.
\clearpage
\begin{figure}
\centering
\includegraphics[width=0.9\textwidth]{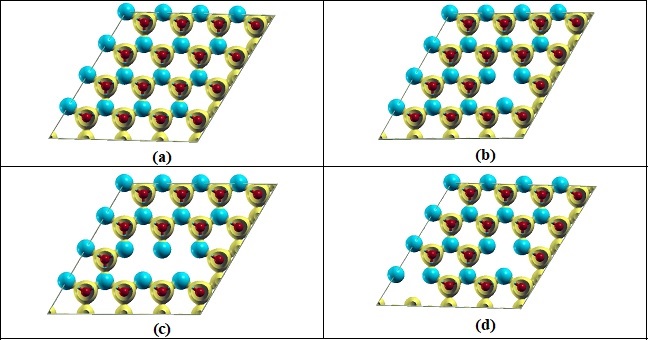}
\caption{Electron charge density plots of MgO (111) monolayers (a) P (b) $V_{O}$ (c) $V_{2O}$-I (d) $V_{2O}$–II.}\label{Figure_3}
\end{figure}
Figure 3 illustrates the electron charge density plots of considered MgO monolayers.  The electron density in pristine MgO ML is delocalized over the Mg-O bridge regions, which gives rise to covalent nature of Mg-O bonds due to the sp2 hybridization.  Such covalent nature of pristine MgO monolayer has been predicted by many theoretical studies [18, 21, 22 24].  However, the electron density is also slightly inclined towards O atoms due to their higher electronegative nature over Mg atoms (Figure 3(a)).  This is an indication of the presence of some ionic nature of Mg-O bond.  Therefore, the Mg-O bonds in the pristine MgO monolayer can be described to have ionic nature on the top of their covalent characters, which is in line with previous reports [24].  From Figures 3(b) - 3(d), it has been observed that the electron charge distributions of defective monolayers remain unaltered from that of the pristine MgO ML.  This means that creation of oxygen vacancies does not change the covalent nature of the Mg-O bonds.
\subsection {Optical properties}
The optical response of the MgO monolayers towards electromagnetic radiation has been investigated using the HSE functional.  In order to achieve this, we have computed various optical parameters for the studied MgO monolayers.  At first, the real $(\epsilon_{r}(\omega))$  and imaginary parts $(\epsilon_{i}(\omega))$ of the complex dielectric function are measured, which basically portray the reflective and absorptive character of a system, respectively.  The values of $(\epsilon_{r}(\omega))$  and $(\epsilon_{i}(\omega))$ are then utilized to calculate the refractive index $(n(\omega))$, extinction coefficient $(k(\omega))$, reflectivity $(R(\omega))$, absorption coefficient $(\alpha(\omega))$  and electron energy loss function $(L(\omega))$ using the relations (5), (6), (7), (8) and (9), respectively.  The geometrical symmetry of the MgO monolayers lead them to behave isotropically along x and y directions for all optical parameters with $V_{2O}$-I being an exception.  Similar to$V_{2O}$-I, anisotropic optical behaviour has also been noticed for BeO nanosheet having two consecutive divacancies [34].  Therefore, we have calculated all the optical quantities of the isotropic monolayers corresponding to the incident light (electromagnetic radiation) polarized along x-($E\parallel{x}$) and z-($E\parallel{z}$) axes.  The variations of $(\epsilon_{r}(\omega))$ and $(\epsilon_{i}(\omega))$  for an isotropic $V_{2O}$-I along y-direction are included in Figure S3.  The variations of the computed optical parameters of the MgO monolayers in both $E\parallel{x}$ and $E\parallel{z}$ directions are shown in Figure 4 and the deduced important optical properties are listed in Table 2.
\clearpage
\begin{table}
\begin{center}
\caption {The important optical properties of pristine and defective MgO (111) monolayers for  $E\parallel x$ and  $E\parallel z$polarizations.  The photon energy values (in eV) corresponding to the maximum value of the optical quantities are given in parentheses}
\vspace{.51cm}
\begin{tabular} {@{}llll@{}}
\includegraphics[width=.9\textwidth]{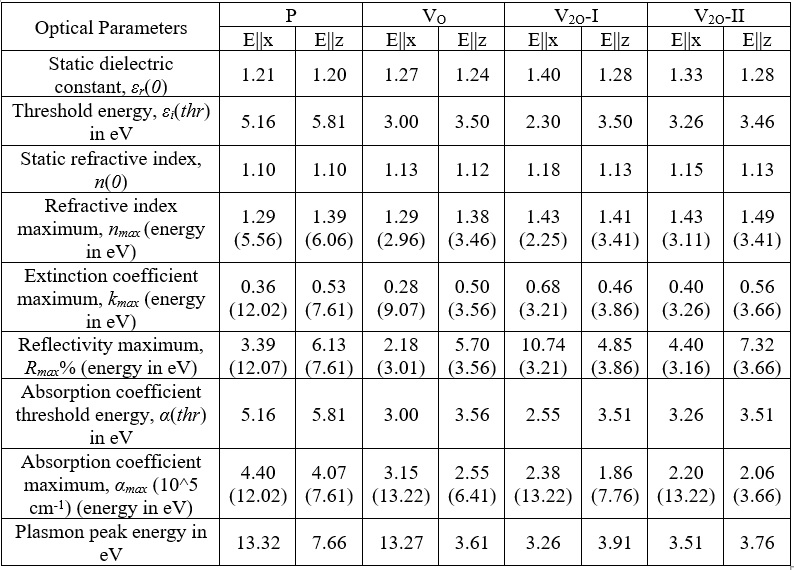} 

\end{tabular}
\end{center}
\end{table}
\clearpage
\begin{figure}
\centering
\includegraphics[width=0.8\textwidth]{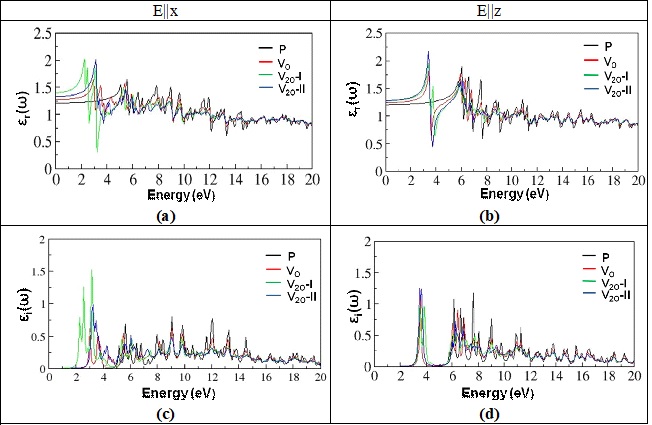}
\caption{Real $(\epsilon_{r}(\omega))$ and imaginary $(\epsilon_{i}(\omega))$  parts of the dielectric function of the MgO (111) monolayers.}\label{Figure_4}
\end{figure}

The values of static dielectric constant $(\epsilon_{r}(0))$, the dielectric constant at zero frequency limit (Figures 4(a) and 4(b)) for P for both $(E\parallel{x})$ and $(E\parallel{z})$ match well with a those of a recent report \cite{bib24}.  The pristine $(\epsilon_{r}(0))$ value calculated earlier at PAW level are found to be slightly greater than the HSE values \cite{bib23,bib24}.  When oxygen atom vacancies are introduced into the MgO monolayer, the $(\epsilon_{r}(0))$ values increase from that of P following the trend $P< V_{O} < V_{2O}–II < V_{2O}$-I for $ (E\parallel{x})$ and $P< V_{O}< V_{2O}–II=V_{2O}-I$ for $(E\parallel{z})$.  The trend observed for $E\parallel x$satisfies the relation  $\epsilon_r(0)=1+(\frac{h\omega_p}{2\pi E_g})^2$given by Penn Model \cite{bib55}, which reflects the inverse relation between $\epsilon_r(0)$ and energy gap (Eg).  The deviation from Penn Model along the $E\parallel z$ direction is due to the discontinuity of the MgO monolayer along z-direction.  Figures 4(c) and 4(d) present the variations of $\epsilon_{i}(\omega)$ of all the considered monolayers.  For P monolayer, the first peak of $\epsilon_{i}(\omega)$ is due to the interband electronic transitions from VBM (O (2p) states) to CBM (Mg (3s) states), which can be inferred from the DOS plots in Figure 2.  For the defective monolayers, $V_{O}$, $V_{2O}$-I and $V_{2O}$ –II, the first peaks are observed at relatively lower energies, i.e., almost near the visible range as shown in Table 2 (2.55-3.56 eV).  Several types of electronic transitions may be responsible for these low energy peaks, such asfrom the VBM (O (2p)) to defect levels, from defect levels to CBM (Mg (3s)) and in between the defect levels as seen in the DOS plots.  The threshold energy values of imaginary dielectric function of the considered monolayers (Table 2) signify the optical gap \cite{bib23}.  The optical gap for  polarization is slightly lower than that of $E\parallel z$ polarization, which agrees well with previous theoretical results \cite{bib23}.  The optical gap values of the MgO monolayers are in consistent with the computed electronic band gap values discussed in section 3.2.

The calculated $n(\omega)$ values for both polarizations are shown in Figures 5(a) and 5(b).  The patterns of these refractive index curves are similar to those of $(\epsilon_{r}(\omega))$.  The static refractive index n(0) follows the same trend as that of $(\epsilon_{r}(0))$ for both polarizations.  It has also been observed that the n(0) values of P are similar to the HSE06 level computed values available in a very recent literature \cite{bib24}.  It is well known that large values of  $n(\omega)$ correspond to strong interaction of incident photons with valence electrons, which lowers the speed of photons during transmission \cite{bib24}.  For systems with vacancies, the maximum peak positions of  $n(\omega)$ get shifted towards lower energies compared to that of P.  With increment in energy, $n(\omega)$ in both polarization directions are found to increase till nmax values (Table 2) are reached (Figures 5(a) and 5(b)).  The nmax values of $V_{O}$ and $V_{2O}$-I monolayers along $ (E\parallel{x})$ are observed in the visible region, while they appear in the UV region for all other cases. The energies at which extinction coefficient maxima $(k_{max})$ appear signify the fastest and the strongest absorption of photons \cite{bib23}.  Similar to  $n(\omega)$, red shift of kmax energies (Table 2) is noticed for defective systems with respect to the pristine monolayer.  The kmax for all the systems are in the UV range in both $ (E\parallel{x})$and $ (E\parallel{z})$ polarizations.  However, it is also noticeable from Figure 5(c) that  $k(\omega)$ spectrum of $V_{2O}-I$ for $ (E\parallel{x})$ direction exhibits prominent peaks even in the visible region, where the nmax of the monolayer lies.  Therefore, it is clear that interaction of incident photons as well as their higher absorption is possible in the visible range for $V_{2O}-I$  monolayer of MgO.  The reflectivity spectra are shown in Figures 5(e) and 5(f) for both$ (E\parallel{x})$ and $ (E\parallel{z})$ polarizations.  The reflectivity values are less than 10\% for pristine monolayer in both directions, which agree with earlier reports \cite{bib24}.  Such low values of reflectivity are also shown by the defective monolayers, which however is remarkably higher for $V_{2O}$-I along $ (E\parallel{x})$ with reflectivity of 10.74\% (Table 2).  The highest values of reflectivity for all the MgO monolayers along with their entire reflectivity pattern cover mostly the UV region of the electromagnetic spectrum.  In exception, the reflectivity pattern covers the range from visible to near UV region in $V_{2O}$-I along $ (E\parallel{x})$ direction as seen in Figure 5(e).
\begin{figure}
\centering
\includegraphics[width=0.8\textwidth]{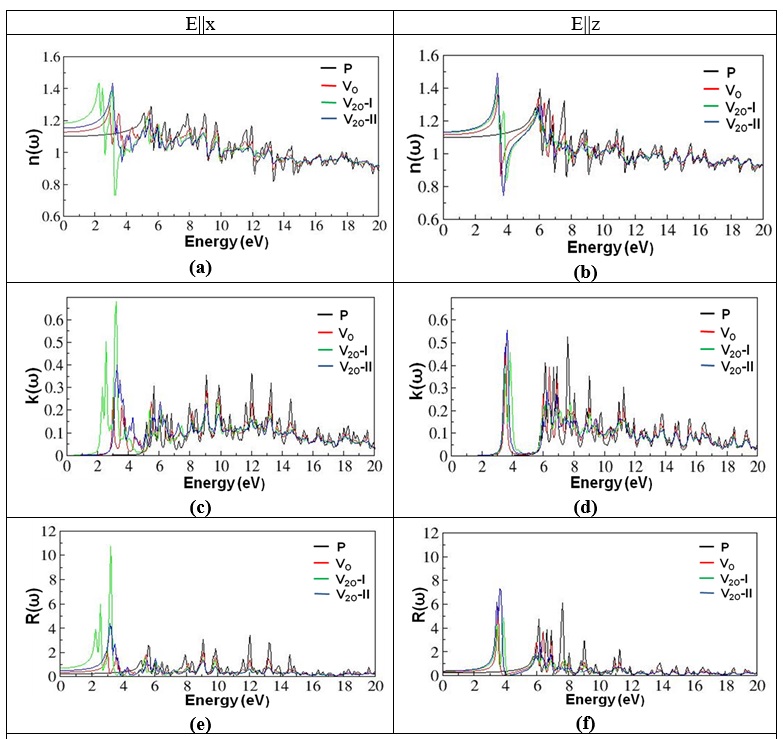}
\caption{Refractive index $(n(\omega))$, Extinction coefficient $(k(\omega))$ and Reflectivity $(R(\omega))$ of the MgO(111) monolayers along $E\parallel x$ and $E\parallel z$ directions.}\label{Figure_5}
\end{figure}
\clearpage
\begin{figure}
\centering
\includegraphics[width=0.8\textwidth]{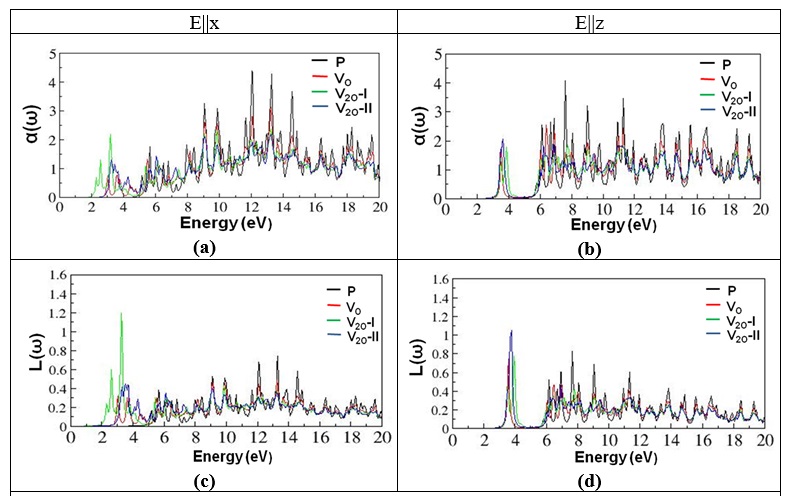}
\caption{ Absorption coefficient $(\alpha(\omega))$ and Energy loss spectrum $(L(\omega))$ of the MgO (111) monolayers along $E\parallel x$ and $E\parallel z$ directions.}\label{Figure_6}
\end{figure}

The calculated absorption coefficient $\alpha(\omega)$  is presented in Figures 6(a) and 6(b) for $E\parallel x$ and $E\parallel z$polarizations, respectively.  The absorption thresholds for the pristine and defective MgO monolayers are located near the energy gaps.  Due to the introduction of oxygen vacancy defects pristine MgO monolayers start absorbing light of lower energies in the near ultraviolet regime, which make them more potential candidates for optoelectronic applications.  On the other hand, the absorption of $V_{2O}-I$ for $E\parallel x$ takes place in a much wider range starting from visible to UV regions (Figure 6(a)) revealing its enhancement in optical absorption behaviour compared to the other systems.  Such improvement in absorption has also been observed for BeO monolayer with double oxygen vacancies \cite{bib34}.  All the 2D systems have their absorption coefficient maxima in the ultraviolet region (Table 2).  The energy loss function $L(\omega)$  represents the energy loss of an electron when travelling through a medium.  The sharp peaks in $L(\omega)$ spectra denote plasmon resonance peaks arising out of collective excitations of electronic charge density of a material.  The frequencies corresponding to the plasmon peaks are known as plasmon frequency.  In Table 2, the plasmon peak frequencies (energies) are found to occupy the UV part of the spectrum.  Red shift of these plasmon peaks is observed for the defective MgO monolayers because of the narrowing of band gaps due to the presence of defect states.  The  $L(\omega)$  spectra of defective MgO monolayers cover a wider range in electromagnetic spectra than that of pristine as observed for other optical parameters.  It is clear from the figure 6(c) that $V_{2O}-I$ suffers electron energy loss even in the visible range, which is revealed by the presence of a prominent peak at energy of 2.60 eV.
	On the basis of the analyses of various optical parameters, we have found that incorporation of oxygen vacancies not only improves the optical response of MgO monolayer but also enhances the operation spectral range.  Interestingly, the defective MgO monolayer with consecutive double oxygen vacancy sites shows the best optical absorption behaviour in a much wider spectral range from visible to near UV region for$E\parallel x$ polarization.

\section {Conclusion}
In the present work, DFT calculations have been performed to study the electronic and optical properties of MgO monolayers including oxygen vacancy defects.  Accordingly, we have chosen oxygen vacancy concentration of 6.25\% and 12.5\% corresponding to one $(V_O)$ and two vacancy sites $(V_{2O})$, respectively, in a 4x4x1 MgO supercell model.  Five different possible divacancy systems have been optimized, which has resulted in almost similar values of formation energies with slightly different value for two consecutive O missing sites $(V_{2O}-I)$.  Therefore, this particular system along with another $(V_{2O}-II)$ from the remaining four $V_{2O}$ monolayers having two non-consecutive O vacancy positions have been selected for our optical study.  All the MgO monolayers with and without defects are found to exhibit planar structures with no magnetic behaviour.  The calculations carried out at hybrid HSE level have revealed similar electronic band structure for pristine MgO monolayer as reported in some previous research studies.  Pristine MgO monolayer possesses indirect and wide semiconducting electronic band gap of 4.84 eV, which decreases to 2.97 eV (indirect), 2.28 eV (direct) and 2.87 eV (indirect) in $V_O, V_{2O}-I$ and $V_{2O}-II$, respectively, due to the appearance of defect levels.  The relative positions of defect states and the Fermi levels unveil n-type behaviour of the defective monolayers.  It is apparent from the PDOS analyses that Mg (3s) and O (2p) states contribute to the CBM and the VBM of pristine MgO, respectively.  The Mg (3s) and O (2p) states constitute the defect levels that arise in the band gap of defective monolayers.  The delocalization of electron density over the bridge positions of Mg-O bonds imparts their covalent nature, which is retained even after the creation of the oxygen defects.  The reduced band gaps in the defective MgO monolayers enlarge the static values of dielectric constant $(\epsilon_r(0))$ and refractive index (n(0)) along $E\parallel x$ polarization for all the considered defective MgO monolayers compared to those of pristine monolayer.  This facilitates stronger photon-electron interaction in the monolayers.  The threshold energies of imaginary dielectric constant (along $E\parallel x$) representing the optical gaps are found to agree well with the electronic band gaps for the monolayers.  The spectra of refractive index $(n(\omega))$ and extinction coefficient (k$(\omega)$) also signify improvement in the spectral range for optical response of the defective MgO monolayers.  For such systems, red shift in energy to near UV region is observed relative to the pristine monolayer.  Exceptionally, $V_{2O}-I$ exhibits significant optical absorption behaviour in the entire range from visible to UV region of the electromagnetic spectrum.  Reflectivity of our computed MgO monolayers are found to be less $(< 10\%) $in both x and y directions with betterment in reflective power for $V_{2O}-I$ to 10.74\% in $E\parallel x$.  The calculated absorption coefficient ($\alpha(\omega)$) and energy loss function ($L(\omega)$) also suggest enhancement in the range of light absorption ability in the defective MgO monolayers over the pristine and $V_{2O}-I$ rendering the maximum coverage from visible to UV range.  Therefore on the basis of the discussed optical properties, we can suggest MgO monolayers with induced oxygen vacancy defects to be promising materials for optoelectronic applications with the consecutive double vacancy sites being the most suitable option.
\paragraph {}
{\bf Acknowledgments}\\
RH thanks Dibrugarh University for financial support.
\paragraph {}
{\bf Conflict of interest}\\
There is no conflict of interest.

\end{document}